# Broadband mid-infrared perfect absorber using fractal Gosper curve

Jihua Zou,[1,2] Peng Yu,[1,5,*] Wenhao Wang,[1] Xin Tong,[1] Le Chang,[1] Cuo Wu,[1] Wen Du,[1] Haining Ji,[3] Yongjun Huang,[4] Xiaobin Niu,[3] Alexander O. Govorov,[1] Jiang Wu,[1,6,*] and Zhiming Wang[1,7,*]

[1] Institute of Fundamental and Frontier Sciences, University of Electronic Science and Technology of China, Chengdu 610054, China

[2] School of Electronic Science and Engineering, University of Electronic Science and Technology of China, Chengdu 611731, China

[3] School of Materials and Energy, University of Electronic Science and Technology of China, Chengdu 610054, China

[4] School of Information and Communication Engineering, University of Electronic Science and Technology of China, Chengdu 611731, China

[5] peng.yu@uestc.edu.cn
[6] jiangwu@uestc.edu.cn
[7] zhmwang@uestc.edu.cn

**Abstract:** Designing broadband metamaterial perfect absorbers is challenging due to the intrinsically narrow bandwidth of surface plasmon resonances. Here, the paper reports an ultra-broadband metamaterial absorber by using space filling Gosper curve. The optimized result shows an average absorptivity of 95.78% from 2.64 to 9.79 μm across the entire mid-infrared region. Meanwhile, the absorber shows insensitivity to the polarization angle and the incident angle of the incident light. The underlying physical principles, used in our broadband absorber, involve a fractal geometry with multiple scales and a dissipative plasmonic crystal. The broadband perfect absorption can be attributed to multiple electric resonances at different wavelengths supported by a few segments in the defined Gosper curve.

## 1. Introduction

Metamaterials have intriguing electromagnetic properties with tailored permittivity and permeability that are not available in nature, promising for perfect lenses, holographic display and invisible cloaking [1, 2]. One important application based on metamaterial is absorber with decreasing dimensions and independent of polarization and incidence angle compared with their conventional counterparts [3-5]. As an important branch of optical and optoelectronic devices, metamaterial absorbers (MMAs) demonstrated a variety of applications at different frequencies [5], such as thermal imaging [6-9], thermal emitters [6, 10], solar photovoltaics [11-13], sensors [14], absorption filtering [15], bolometers [16], and single photon source [17].

MMAs can be divided into two types—narrow band and broadband [4, 5, 18]. Narrow band MMAs have been demonstrated extensively, such as single-band [3, 6, 8, 19, 20], dual-band [10, 21-23], triple-band [24, 25] and multi-band [26-28]. However, achieving broadband MMAs at a longer wavelength—they are indispensable for thermophotovoltaics, photodetection, bolometry, radar stealth, manipulation of mechanical resonances—still remains a challenge due to the intrinsically narrow bandwidth of localized surface plasmon resonances (LSPRs) or surface plasmon polaritons (SPPs) generated on metallic nanostructure [5]. Broadband mid-infrared (MIR) MMAs are appealing to various applications, e.g. thermal imaging system with spatial light manipulation [9], thermophotovoltaics with efficiency above SQ limit [29], gas or molecular sensing with low cost [30]. A widely used approach is integrating multi-shaped or multi-sized resonators in a unit cell, which broadens absorption bandwidth by merging several adjacent modes [31]. However, the bandwidth, in particular the band width of 90% absorption, can only be increased to some extent because of the limited number of resonators in a unit cell. Alternatively, vertically stacking resonators can also improve the bandwidth, but fabrication with aligned nanoscale features is time-consuming [32, 33]. By exciting LSPR of nanoparticles, nanocomposite-based MPAs can give rise to broadband absorption in UV-NIR spectral region, without introducing photolithography and electron beam lithography [34, 35]. However, the absorption band is limited in the UV-NIR wavelengths due to plasmonic resonances of metallic nanoparticles.

MMAs based space-filling design offers a compelling alternative with a facile and straightforward nanofabrication—typically, the size and shape of resonators are controlled by a certain algorithm [5]. These structures adopt a single noble metallic layer and a ground layer separated by dielectric spacer to form a resonant cavity that capture light [36]. Chaos pattern [37], Peano and Hilbert curve [38] were shown for broadband absorption in GHz region. In the MIR region, T. Mayer et al. demonstrated a MMAs with absorption over 98% between 1.77 and 4.81 μm by using a genetic algorithm [36]. Gao et al. proposed a MMA with binary-pattern which demonstrated absorption over 95% from 3 to 3.9 μm. However, achieving perfect absorption in the whole MIR (3-8 μm) has so far remained a challenging task.

Here, based on the space-filling Gosper curve, we propose a broadband MIR MMAs that demonstrates a nearly-perfect absorption covering the entire MIR—over 95% absorption spanning from 2.6 to 9.8 μm. Gosper curve is a space-filling curve that can be represented using a L-system with rules shown in [39]. We simulated first three stages in the construction of a fractal array based on Peano-Gosper space-filling curve (Figs. 1(a)-1(c), i.e. P = 1, P = 2, and P = 3). In order to minimize reflection and maximize absorption in the MIR region, a dielectric layer (n=1.5) is used to tailor the impedance match to free space; platinum (Pt) is used to replace conventional gold resonator to increase the bandwidth of MMAs (see Appendix A, Fig. 7). It was shown in the past that a fractal geometry (plasmonic Cayley-tree nanostructure made of gold in [40].) can give a broadband spectrum with multiple sharp peaks. However, we demonstrate here that it is crucial to use a plasmonic crystal with strong damping (like Pt) to realize a spectrum with continuous and very strong absorption in a broadband spectral interval. Simulations show an angle-independent absorption greater than 90% in $\pm 60°$ range under either transverse electric (TE) or transverse magnetic (TM) polarization. To the best of our knowledge, the MMAs based on Peano-Gosper space-filling curve has a highest absorption preserved a broadest bandwidth in the MIR region among its single layer counterparts. We believe the proposed device will spark applications of MMAs in nanophotonics and optoelectronics, e.g. thermal imaging, biomedical sensing, and free-space communication. For example, our device may have applications on surface-enhanced infrared spectroscopy—infrared vibrations of molecules located in these fields can be boosted by orders of magnitude and permitting a spectroscopic characterization with unprecedented sensitivity.

## 2. Method

The finite-difference time-domain (FDTD) and Finite Element Method (FEM) were employed to simulate the MMAs from Lumerical and COMSOL Multiphysics, respectively (see Appendix B, Fig. 8). In the *z* direction, we used perfectly matched layers (PML) boundary conditions to eliminate the boundary scattering. A single unit of the MMAs is simulated with periodic boundary in the *x* and *y* directions and the periodicity is defined as *p* (as shown in Fig. 1, all the illuminations are x polarized except Fig. 6). The thickness of the ground metallic layer is fixed at 150 nm. Absorption (*A*) was calculated by $A(\lambda) = 1 - R(\lambda) - T(\lambda)$, where $A(\lambda)$, $T(\lambda)$ and $R(\lambda)$ are wavelength-dependent absorption, transmission and reflection, respectively. The metallic ground plane is several times the skin depth for Pt and it blocks the incident light. Therefore, the absorption (*A*) can be simplified as $A(\lambda) = 1 - R(\lambda)$.

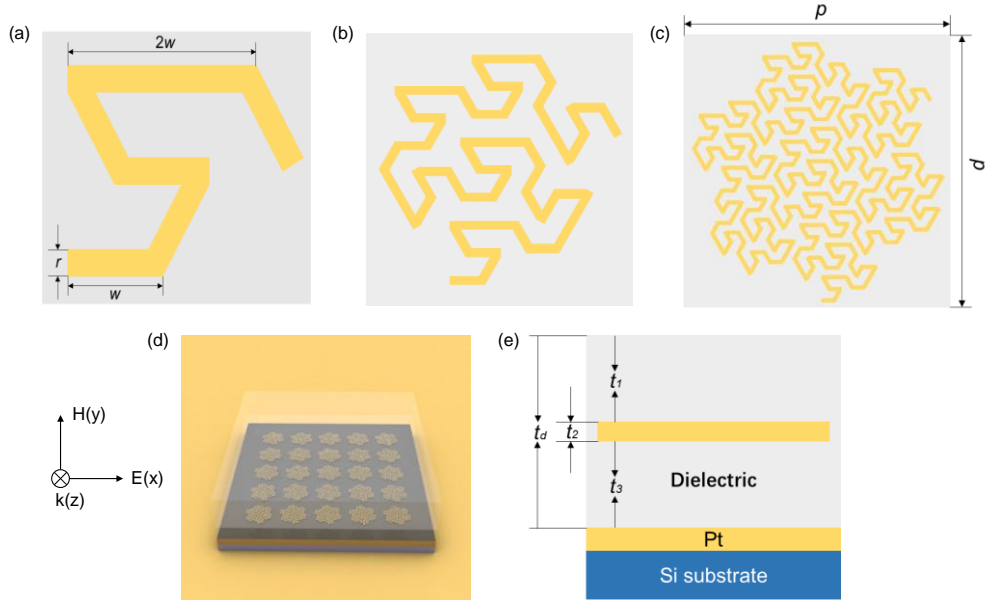

Fig. 1. Schematic of the MMA based on Peano-Gosper space-filling curve. (a)-(c) show the stage 1, stage 2 and stage 3 evolution of the Peano-Gosper curve. (d) A schematic diagram of the proposed MMAs. (e) Cross-sectional view of the four-layer Peano-Gosper based MMAs.

## 3. Results and discussion

### 3.1. Performances and principles

The resonators embedded in the dielectric layer are derived from the Peano-Gosper space-filling curve (Fig. 1) [38]. The stage 1, stage 2 and stage 3 evolutions of the Gosper curve are shown in Figs. 1(a)-1(c). In this paper, the higher-stage Gosper curves are implemented by the rotation of a stage 1 Gosper curve, i.e. $\pm 120°$. When we give this curve a certain length ($w$, $2w$), a certain width($r$) and a certain height ($t_2$), it becomes the nanostructure shown in Fig. 1(c). In this paper, we only focus on the stage 3 MMAs because it provides a broad bandwidth in the NIR with intense absorption when compared with stage 1 and 2 due to resonances from the different parts of the resonators at specific wavelengths shown in Fig. 2(a). The schematic of a stage 3 MMAs is shown in Figs. 1(d) and 1(e). The thickness of the dielectric layer ($t_d$) is the sum of $t_1$, $t_2$ and $t_3$. The structure is composed of four layers, including a bottom metallic layer, a dielectric spacer, the Peano-Gosper resonator and a top dielectric layer. Broadband light consumption in these layers can be explained as follows: the bottom three layers form a resonant EM cavity, generating multiple overlapping resonances in the MIR range; the top layer provides an impedance match between Peano-Gosper resonator and free space; moreover, transmitted light by the top three layers can be further consumed in the metal ground plane [38]. The resonator is critical in design MMAs [5]. The Peano-Gosper shape provides an excellent framework for designing broadband planar resonators. The multiple segments with in the Peano-Gosper shape have different orientations and lengths that can support multiple resonances in the MIR range. To highlight the advantages of the Gosper MMAs, the bandwidth and average absorption of the Gosper MMAs is compared with the state-of-the-arts of the MIR MMAs (Table 1). Compared with conventional ways of adjusting the size and shape of a resonator, our methodology adds another dimension in designing complex resonators and shows clear advantages in both bandwidth and absorption for MIR perfect absorbers.

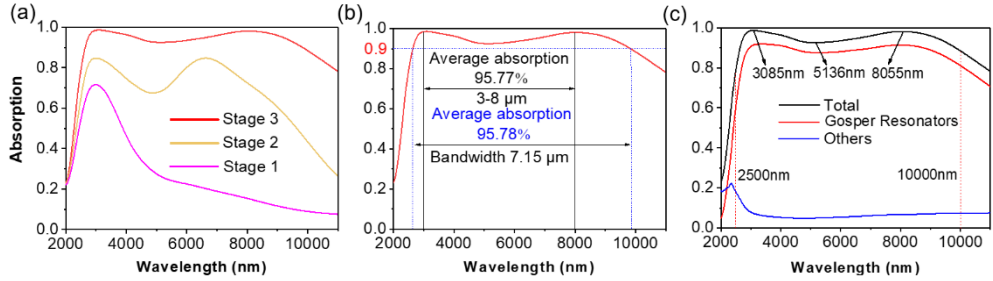

Fig. 2. (a) Absorption spectra of the stage 1, stage 2 and stage 3 of the MMAs. (b) Simulated absorption spectrum of the stage MMAs with average absorption and bandwidth shown. (c) Absorption contribution from the Gosper resonator and the ground layer of each part of the MMAs. Here, thickness of the top dielectric layer ($t_1$), nanostructure ($t_2$), and the bottom dielectric ($t_3$) are 595 nm, 60 nm, and 595 nm, respectively. The length ($w$) and width ($r$) of a single cuboid is 80 nm and 40 nm, the unit size ($p$) is 230 nm, 630 nm and 1550 nm, respectively.

**Table 1.  Selected Publications on Absorbers Operating Across the MIR Region.**

| Ref. | Average absorption(%) | Bandwidth (μm) | Range (μm) |
| --- | --- | --- | --- |
| This paper | 95.8 | 7.15 | 2.64-9.79 |
| [41] | 94.5 | 3.17 | 4.27-7.44 |
| [42] | 95.0 | 2.00 | 8.50-10.50 |
| [36] | 98.0 | 3.04 | 1.77-4.81 |
| [32] | 97.5 | 2.8 | 3.00-5.80 |
| [43] | 96.0 | 0.46 | 3.19-3.65 |
| [44] | 98.0 | 0.30 | 3.60-3.90 |
| [37] | 90.0 | 2.30 | 1.9-4.2 |
| [45] | 70.0 | 3.0 | 4.0-7.0 |

The absorption spectra of the stage 1, stage 2 and stage 3 at normal illumination are shown in Fig. 2(a). From the stage 1 to stage 3, the absorption and bandwidth of MMAs are significantly boosted. As the stage increases, the space of adjoining elements narrows and the number of resonators increases. The coupling increases, which leads to enhanced absorption. We further analyze the absorption at stage 3 as shown in Fig. 2(b). It shows that over 90% of the incident light is efficiently absorbed (an average absorption of 95.78%) across the MIR spectrum from 2.64 to 9.79 μm (7.15 μm bandwidth). The perfect light absorption in MMAs is caused by LSPR usually takes place near the Gosper resonator, while the loss due to dielectric nature of Pt should reside in the ground metallic layer. We calculate the absorption contribution of the elements by $P_{abs} = (1/2)\omega\varepsilon''|E|^2$, where ω is the angular frequency, ε" is the imaginary part of the permittivity, and |$E$| is the amplitude of the total electric field confined in the material, as shown in Fig. 2(c). For wavelengths below 3000 nm, the EM energy loss is roughly equally distributed in Gosper resonators and the ground metallic layer, leading to a low absorption. In fact, a Pt film itself can be a good absorber at short wavelengths and thereby it possesses a decent absorption coefficient without any plasmonic resonances. However, in the range of $\lambda$ > 3000 nm, the single Pt film can no longer support good absorption and the absorption is contributed primarily to the plasmonic resonances. Meaning that the EM fields penetrate into the ground layer decrease significantly and the light is harvested by the top Gosper resonator (see Appendix C, Fig. 9).

In order to understand the absorption mechanism of the proposed MMAs, we studied the electric field /$E$/ in the *x-z* plane, magnetic field /$H$/ in the *y-z* plane, and power of absorbed light $P_{abs}$ in the *x-z* plane. The MMAs are illuminated by the normal incidence of light at various resonant wavelengths (i.e. 2500, 3085, 5136, 8055 and 10000 nm**,** Fig. 3). It shows an increase in the electric field distribution at certain locations of the nanostructuresat different resonance wavelengths. The electric filed is localized near the surface of the Gosper curve and it is further confirmed by electric field profiles along *x-y* plane (see Appendix D, Fig. 10). These field localizations suggest that surface plasmons are excited at these wavelengths, in accordance with LSPR of Pt in the infrared region [46]. Off the resonance at 2500 nm, the electric field enhancement is negligible, while large field enhancements are observed on different segments of the resonator at other wavelengths. The broadband absorption only originates from electric resonance (see Appendix E, Fig. 11) and this was also observed by J. A. Bossard et al. using space filling genetic algorithm curve [36]. The thick dielectric substrate prevents strong coupling between the Gosper resonator and the ground plane.

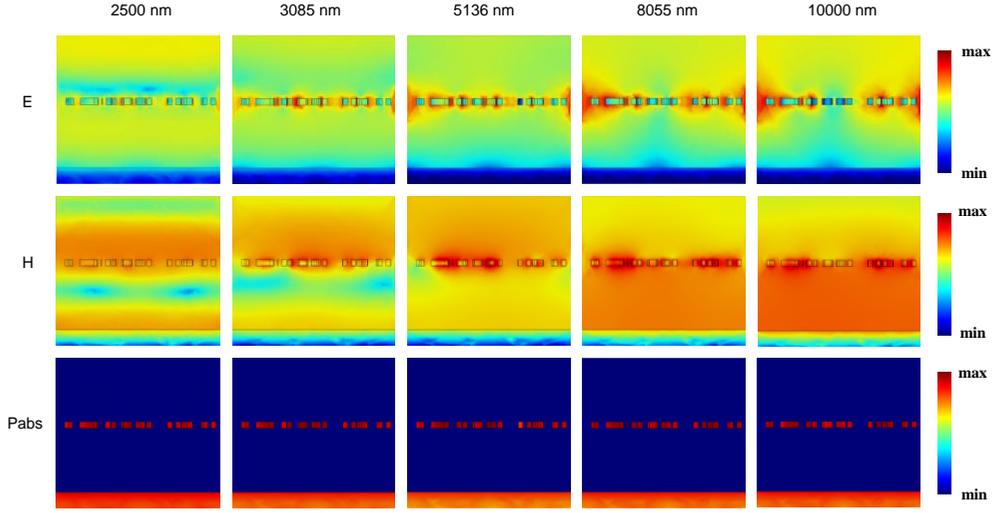

Fig. 3. Cross-sectional view of the relative electric field magnitude |*E*|, magnetic field magnitude |*H*|, and $P_{abs}$ in the unit cell at λ = 2500, 3085, 5136, 8055, and 10000 nm.

The charge and energy flow distributions are shown to understand the broadband absorption in the Gosper MMAs. It shows the charge distribution induced by illumination at different wavelengths [2500, 3085, 5136, 8055 and 10000 nm, Fig. 4(a)]. The accumulated charges, distributing inside the Gosper antennas, form alternating negative charges (−) and positive charges (+), and demonstrate a clear local "dipole" distribution. As the wavelength of incoming photons increases, the charge accumulation clearly increases in different segments of the Gosper layer. Light first propagates downwards but a large amount of light is absorbed and confined in the Gosper resonator as shown by the energy distribution shown in Fig. 4(b). The EM energy locates at where the magnetic field is concentrated. This again justify that the broadband absorption originates from the resonance of electric dipoles.

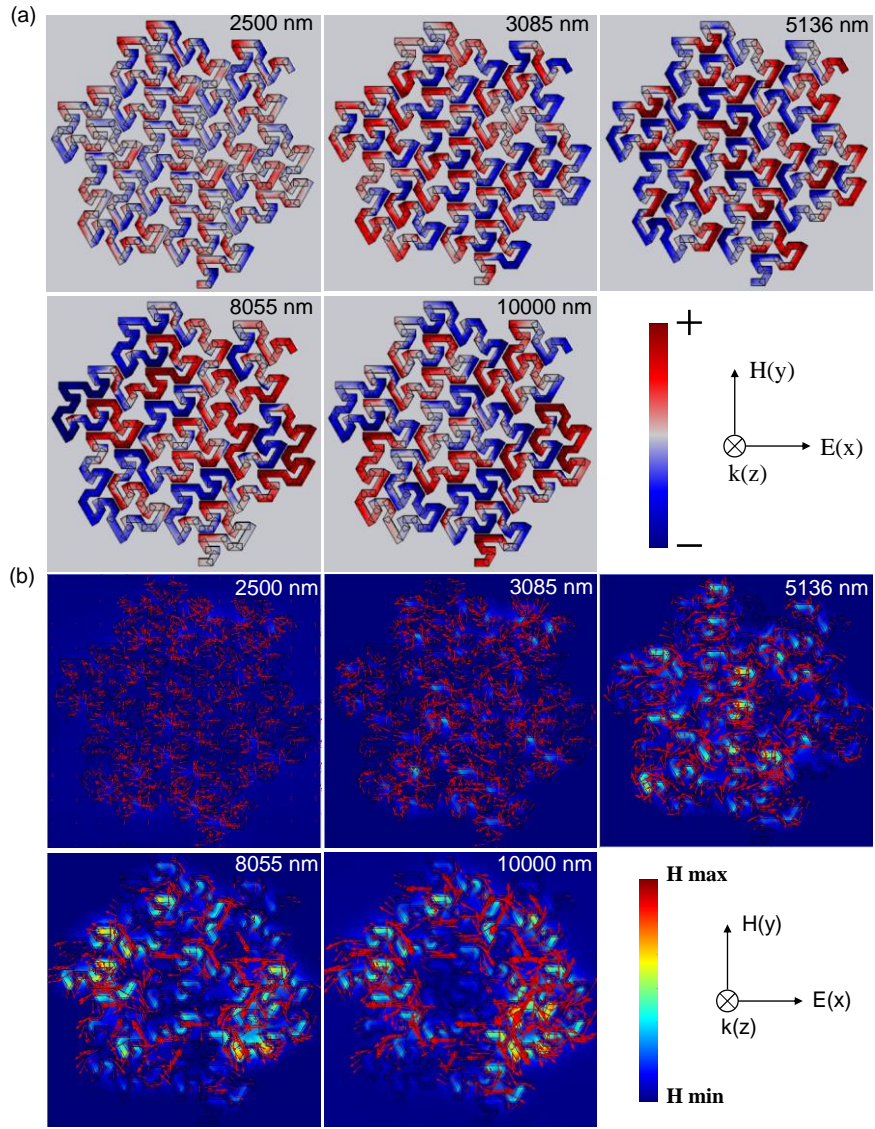

Fig. 4. (a) The charge distributions in the Gosper layer at resonant wavelengths of 2500, 3085, 5136, 8055 and 10000 nm. (b) Magnetic field and energy flow distribution (arrow maps) of Gosper layer at resonance wavelengths of 2500, 3085, 5136, 8055 and 10000 nm.

## 3.2. Effects of structural parameters on absorption

The absorption of the MMAs can be well tuned by the structure of the metamaterial absorber. The relevant results are shown in Fig. 5. Firstly, we investigate the absorption spectrum of the MMAs with different dielectric layer thicknesses ($t_d$), as shown in Fig. 5(a). The absorption peaks shift to red and the bandwidth broadens gradually when the dielectric thickness increases—a perfect absorption spectrum is achieved around $t_d$ =1250 nm. This can be explained by the degree of impedance match to free space [38]. Also, the thickness of the nanostructure layer ($t_2$), the size of a unit cell (*p*), and length (*w*) of the MMAs all play a role in determining the absorption and its bandwidth shown in Figs. 5(b)-5(d). Compared with the unit cell size and length, the thickness ($t_2$) of the resonator has a relatively small influence on the absorption and bandwidth. As the thickness increases, it induces a weaker coupling, but as the thickness increases and it induces a weaker coupling [47]. Thus, the $t_2$ need to be compromised between high absorbance and broad bandwidth. The unit cell size (*p*) is directly linked to the resonance strength of adjacent cells—an optimized bandwidth and absorption achieved at *p*=1550 nm (When *p* is 1530 nm, absorption does not change much and the unit cell is close to the boundary of the simulation region) and with the increase of the period, the resonance of the adjacent periodic units decreases and the absorption tends to decrease gradually. The *w* of the MMAs played a critical role in tailoring the absorption bandwidth. Especially when *w*=70 nm and *r*=40 nm, the bandwidth drops significantly. When *r* remains unchanged and *w* has a small value, nanostructure becomes a thin film and the absorption drops sharply. Conversely, when *w* has a large value, the resonance between adjacent units is weak, and the absorption will also decrease.

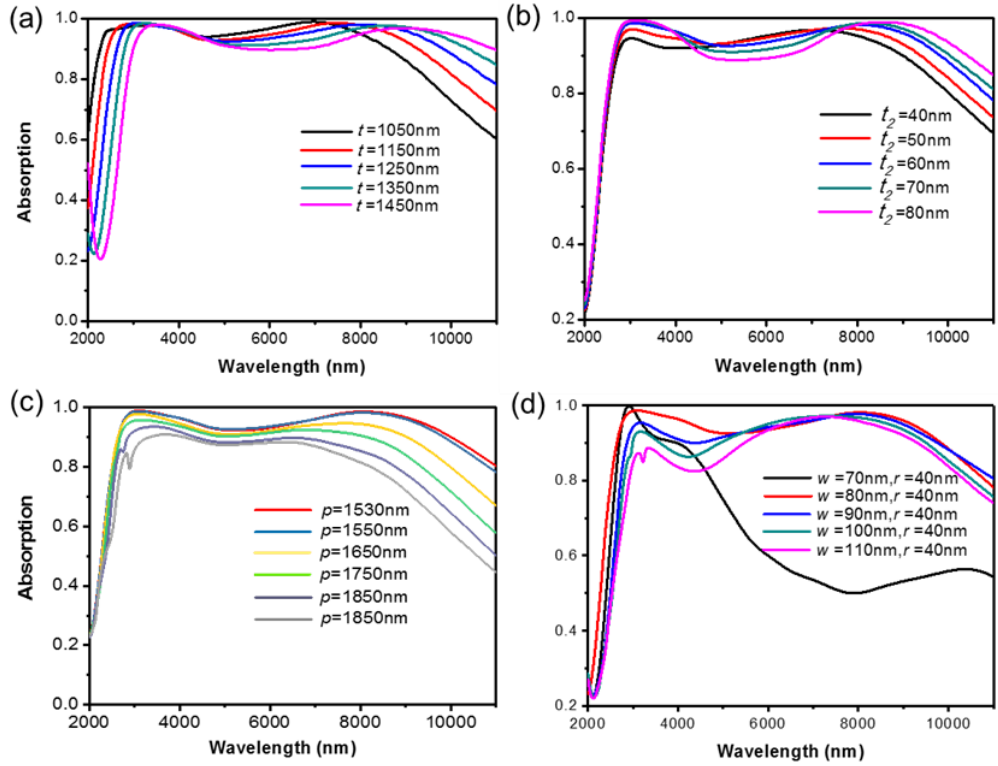

Fig. 5. Influence of the geometric parameters on the absorption performance under the normal

incidence of TM-polarized light. Simulated absorption spectra with different dielectric thicknesses $t_d$ (a), Gosper layer thickness $t_2$ (b), the period $p$ (c), and size $r$, $w$.

## *3.3. Absorption at different polarization angles and incident angles*

We further investigate the absorption spectra of the Gosper MMAs under TE and TM polarization and oblique angles. Due to the asymmetric shape of the Gosper resonator, as the polarization angle of the incident light changes, the average absorption and bandwidth decrease but the average absorption between 3-8 μm maintains above 90% shown in Figs. 6(a) and 6(b). Absorption under oblique incidence for TE and TM polarized illuminations are compared [Figs. 6(c) and 6(d)]. The structure demonstrates almost angle-independent absorptions for TE and TM waves at a relatively small angles (0° to 40°). The angular sensitivity of the absorption for both TE and TM wave can be attributed to sensitivity of surface plasmon [48]. Moreover, the TM absorption at 8 μm drops more quickly with increasing angle when compared with TE wave, while a dip and peak absorption at 5 and 7.5 μm become evident for TE wave at a large incident angle. In a word, the absorber is insensitive to the incident angle. The average absorption from 3 to 8 μm still maintains 84% when the angle reaches up to 60°. This indicates that the Gosper MMAs proposed here is also applicable for devices under constant change of incident angle or/and polarization, i.e. the sun.

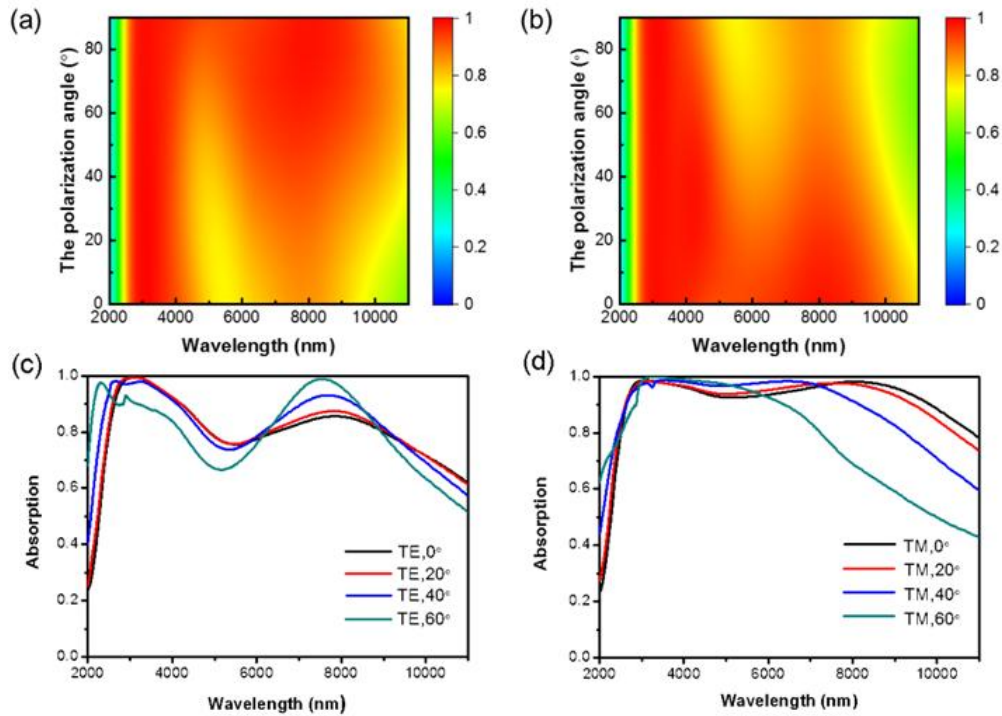

Fig. 6. Contour plot of Gosper MMAs absorption under TE(y-polarized) (a) and TM (x-polarized) (b) illumination with an increasing step of 10 ° from 0 to 89 °; absorption spectra with different Incident angles for (c) TE and (d) TM wave with an increasing step of 20 ° from 0 to 60 °.

## 4. Experimental feasibility

Due to the limitation of experimental conditions, this paper is limited to simulation work. However, it can be easily fabricated by current nanofabrication. Electron beam lithography (EBL) and standard lift-off process can be utilized to fabricate the MMA with Gosper Resonators. The practicable experimental processes are shown here: first a 5 nm Ti adhesive layer and a 150 nm Pt film can be deposited on a Si substrate subsequently by electron-beam evaporation; following that, a 595 nm thick dielectric precursor layer can be formed via spinning coating and then it is further heated and treated; a PMMA layer is coated on the sample for EBL process to define the Gosper patterns; then, a 60 nm Pt is evaporated and followed by a lift-off process to form the Gosper resonators. Finally, a 595 nm thick dielectric layer is deposited atop.

## 5. Conclusion

In this paper, we reported a broadband based on space-filling Gosper curve and it demonstrates the highest bandwidth in the near-mid infrared region with nearly perfect absorption among their single layer counterparts. The proposed MMAs exhibits a nearly perfect absorption with an average absorbance of 95.78% and 90% absorption bandwidth over 7.15 μm from 2.64 to 9.79 μm. Meanwhile, the MMA is insensitive to the incident angle. Especially for TM mode, 90% absorption bandwidth still maintains as high as 3.53 μm at the incident angle of 60 °, and the average absorptivity is 97.64% from 2.87 to 6.40 μm. The broadband and nearly-perfect MIR absorption can be attributed to the defined the different components of the Gosper curve that support multipole resonances at different wavelengths. Interestingly, the electric resonances dominants in the MMAs while its conventional counterparts have electric and magnetic resonances. This work provides a new method for the design broadband absorber and sheds light on practical application, such as infrared imaging, broadband photodetection and biomedical sensing.

## Appendix A

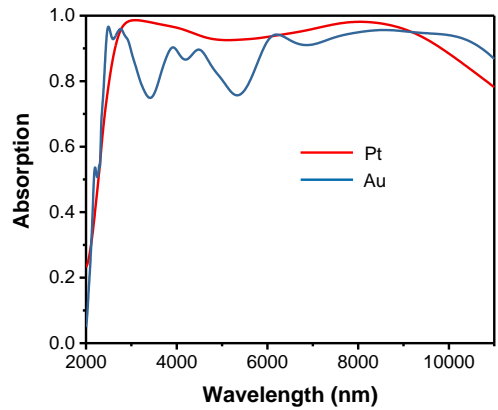

Fig. 7. Absorption of metamaterial absorbers made of Au and Pt.

## Appendix B

For FET simulation, we first used Auto CAD to model the structure, and then imported to Comsol, using the same boundary conditions as FDTD. The refractive indexes of Pt is modelled using fitted optical data from [49], and we set the index of refraction of dielectric to be 1.5.

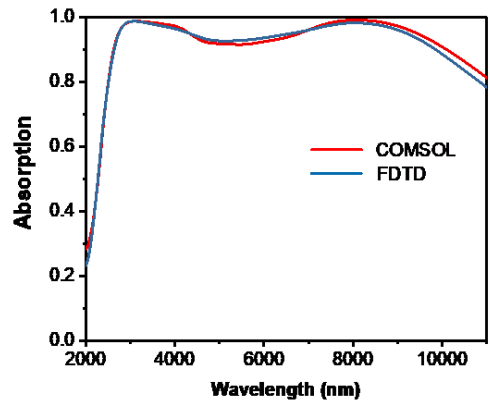

Fig. 8. Duplication of absorption spectra using COMSOL and FDTD.

## Appendix C

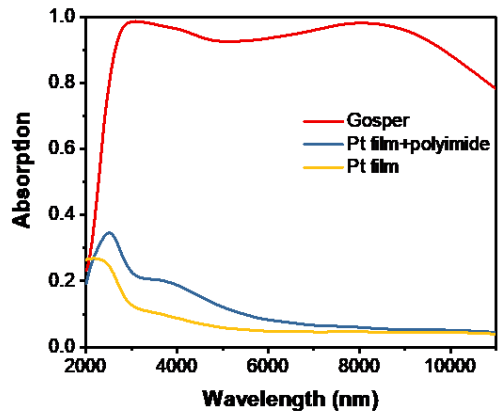

Fig. 9. Simulated absorption of the Pt film only with a thickness of 210 nm (thickness of resonator+ ground film), Pt film in dielectric and MMA.

## Appendix D

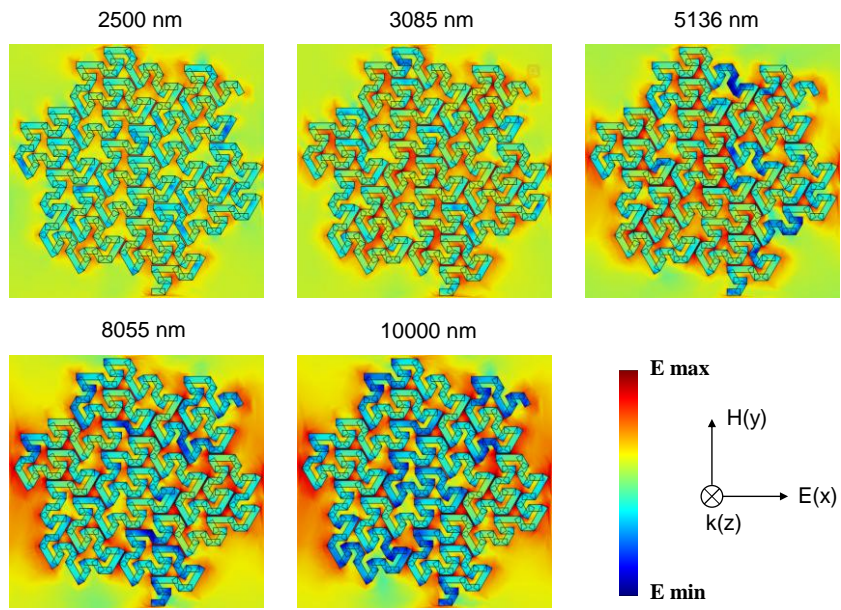

Fig. 10. The electric field distribution (logarithmic scale) in the x-y plane at λ = 2500, 3085, 3085, 5136, 8055, and 10000 nm, respectively.

## Appendix E

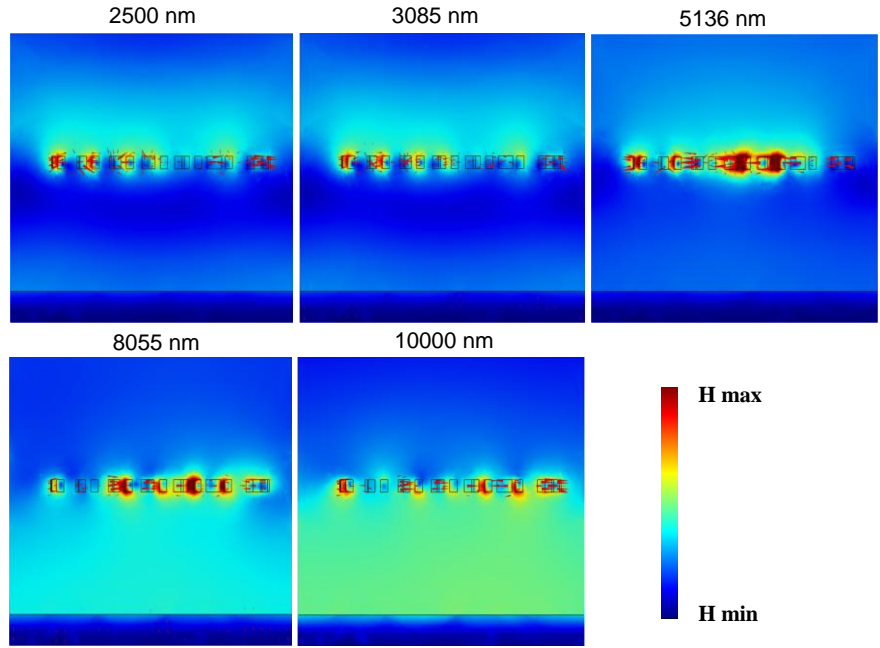

Fig. 11. The magnetic field and current distribution (arrow) of the absorber in the *y-z* plane.

**Funding**

Program 973 (2013CB933301, 2018YFA0306100); National Natural Science Foundation of China (NSFC) (61474015, 51302030).

**Disclosures**

The authors declare no competing interests.